\documentclass{nature}
\usepackage{graphicx}
\makeatletter
\let\saved@includegraphics\includegraphics
\AtBeginDocument{\let\includegraphics\saved@includegraphics}
\renewenvironment*{figure}{\@float{figure}}{\end@float}
\makeatother

\usepackage{multirow}
\usepackage{subfigure}
\usepackage{mathtools}
\usepackage[colorlinks=true,linkcolor=blue,citecolor=black,urlcolor=blue]{hyperref}
\usepackage{authblk}
\usepackage[labelfont=bf]{caption}
\usepackage[figurename=Figure]{caption}
\usepackage{siunitx}
\usepackage[dvipsnames]{xcolor}
\usepackage{soul}
\usepackage{xfrac}

\usepackage{soul,xcolor}
\usepackage{lineno}
\usepackage{siunitx}
\graphicspath{{./figures/}}

\setstcolor{red}

\title{Intrinsic N\'eel antiferromagnetic  multimeronic spin-textures in ultrathin films
}
\author{Amal Aldarawsheh$^{1,2*}$, Moritz Sallermann$^{1,3,4}$, Muayad Abusaa$^5$ and  Samir Lounis$^{1,2*}$}
\begin{document}
\maketitle
\begin{affiliations}
 \item Peter Gr\"{u}nberg Institute and Institute for Advanced Simulation, Forschungszentrum J\"{u}lich and JARA, D-52425 J\"{u}lich, Germany
 \item Faculty of Physics, University of Duisburg-Essen and CENIDE, 47053 Duisburg, Germany
 \item RWTH Aachen University, 52056 Aachen, Germany
 \item Science Institute and Faculty of Physical Sciences, University of Iceland, VR-III, 107 Reykjavík, Iceland
 \item Department of Physics, Arab American University, Jenin, Palestine\
 
 * a.aldarawsheh@fz-juelich.de; s.lounis@fz-juelich.de
 
\end{affiliations}
\section*{Abstract}
 
The realization of topological antiferromagnetic (AFM) solitons in real materials is a major goal towards their use in information technology. While they bear various advantages with respect to their ferromagnetic cousins, their observation is scarce. Utilizing first-principles simulations, here we predict new chiral particles in the realm of AFM topological magnetism, frustrated multimeronic spin-textures hosted by a N\'eel magnetic state, arising in single Mn layers directly grown on Ir(111) surface or interfaced with Pd-based films.  These topological structures are intrinsic, i.e. they form in a single AFM material, can carry distinct topological charges and can combine in various multimeronic sequences with  enhanced stability against external magnetic fields. We envision the  frustrated N\'eel AFM multimerons as exciting highly-sought after AFM solitons having the potential to be utilized in novel spintronic devices hinging on non-synthetic AFM quantum materials.

\section*{Introduction}

Recent experimental breakthroughs promoted antiferromagnetic (AFM) materials into the realm of information technological applications~\cite{Jungwirth2016,gomonay2017concepts,baltz2018antiferromagnetic,gomonay2018antiferromagnetic,vsmejkal2018topological , nvemec2018antiferromagnetic,legrand2020room,Jani2021, aldarawsheh2022emergence} and triggered state-of-the-art activities in the world of topological magnetism~\cite{dohi2019formation,legrand2020room,finco2021imaging,juge2022skyrmions,chen2022controllable}.
The antiparallel spin sublattices present in AFM materials result in zero dipolar fields, making them insensitive to magnetic field perturbations and enhances the stabilization of nanoscale topological structures~\cite{Jungwirth2016,barker2016static,olejnik2018terahertz,gomonay2018antiferromagnetic,dohi2019formation,aldarawsheh2022emergence,aldarawsheh2023}. Moreover, AFM materials possess faster spin dynamics than ferromagnets by orders of magnitude~\cite{kampfrath2011coherent,gomonay2014spintronics,baierl2016terahertz,bhattacharjee2018neel}, which is an appealing characteristic for THz  magnetic memory and logic devices.

The race in identifying AFM  non-trivial spin-swirling objects is strongly motivated by their particle-like nature,  potentially realizing ideal magnetic bits, augmented with the low power consumption~\cite{rosales2015three,barker2016static,zhang2016antiferromagnetic,velkov2016phenomenology,keesman2016skyrmions,jin2016dynamics,gobel2017antiferromagnetic,tomasello2017performance,akosa2018theory,gomonay2018antiferromagnetic,silva2019antiferromagnetic,fernandes2019skyrmions, legrand2020room,gao2020fractional,li2020bimeron,shen2020current,silva2021antiferromagnetic,juge2022skyrmions} involved in their manipulation with the possibility of controlling their current-driven motion~\cite{barker2016static,zhang2016antiferromagnetic,velkov2016phenomenology,jin2016dynamics, gobel2017antiferromagnetic,akosa2018theory,aldarawsheh2022emergence,aldarawsheh2023} while avoiding the skyrmion Hall effect that plagues the ferromagnetic (FM) cousins~\cite{Lin2013,Nagaosa2013,Jiang2016,woo2016observation, Litzius2017}.

\begin{figure}[ht]
	\begin{center}
	\includegraphics[width=1\columnwidth,keepaspectratio]{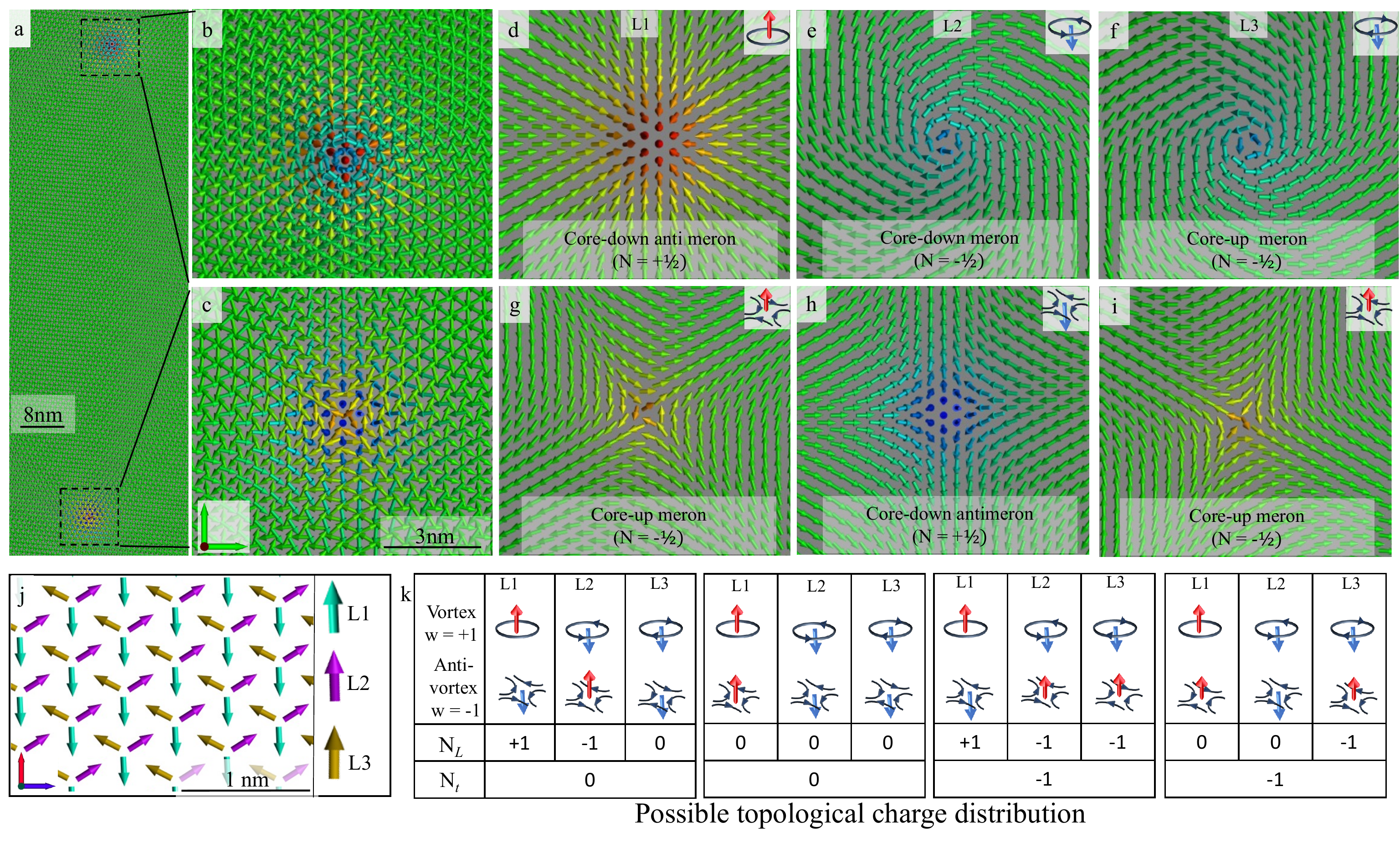}
	\caption{\textbf{Frustrated N\'eel AFM meronic topology}: \textbf{a} AFM hexameronic state composed of Vortex-antivortex pair that emerges in a frustrated triangular Mn layer on e.g. Ir(111) surface  with zoom into the  vortex \textbf{b}, and  antivortex \textbf{c} components. The frustrated AFM meronic texture is decomposed into three FM vortix-antivortex pairs residing at sublattices L1 \textbf{d, g}, L2 \textbf{e, h}, and L3 \textbf{f, i}. \textbf{j} illustration of the N\'eel AFM ground state with colors indicating the decomposition into three sublattices L1, L2 and L3. \textbf{k} Schematic representation of the set of sublattices for the possible topological magnetic structures, with w stands for the winding number and  N stands for  the topological number (N = wp/2 with p = +1 for up and p = -1 for down polarity of the core).}
		\label{fig:1}
	\end{center}
\end{figure}

This led to the recent discovery of synthetic AFM skyrmions,  which consist of two FM skyrmions that are realized in two distinct magnetic layers and antiferromagnetically coupled through a non-magnetic spacer layer~\cite{dohi2019formation,legrand2020room,finco2021imaging,juge2022skyrmions,chen2022controllable}. Here, utilizing first-principles in conjunction with atomistic spin dynamics we unveil multimeronic textures, a new type of topological AFM particles, which are non-synthetic and emerge in magnetically frustrated thin films (see Fig.\ref{fig:1}). 

Regular FM merons are in-plane magnetized textures with magnetization that curls around a stable core pointing out-of-plane, and are topologically equivalent to  one half of a skyrmion~\cite{tretiakov2007vortices,ezawa2011compact,phatak2012direct,lin2015skyrmion,tan2016topology,yu2018transformation,lu2020meron,augustin2021properties,hayami2021meron,xia2022qubits}. The meronic topological charge $N=  \frac{1}{4\pi}  \int   \textbf{n}  \cdot \left ( \frac{\partial\textbf{n}}{\partial_{x}} \times  \frac{\partial\textbf{n}}{\partial_{y}}\right) d{x}d{y}$ equals $\pm\frac{1}{2}$,
where \textbf{n} is the direction vector of magnetization. While they have been observed experimentally in bulk~\cite{donnelly2021experimental}  and  thin films~\cite{yu2018transformation,gao2019creation}, they emerge in AFM synthetic~\cite{phatak2012direct,tan2016topology,kolesnikov2018composite,amin2023antiferromagnetic} and intrinsic bulk (thick films)  phases~\cite{chmiel2018observation,gao2020fractional,jani2021antiferromagnetic}, following a large body of phenomenology-based simulations {~\cite{fernandes2019skyrmions,radaelli2020micromagnetic,li2020bimeron,shen2020current,silva2021antiferromagnetic}.  However, a pristine ultrathin film material that hosts AFM merons remains elusive. 

Our multi-meronic textures are distinct from current predictions since they form in a realistic material a rich set of combinations materializing in a frustrated in-plane N\'eel ground state, shown in Fig.~\ref{fig:1}j, which can be decomposed into three FM sublattices, with an opening angle of 120$^\circ$ between their respective magnetic moments. We predict a single Mn layer as a universal hosting material once interfaced in different fashions with Ir(111) surface with and without Pd, Fe monolayers, PdFe bilayer or Pd$_2$Fe trilayer (see Fig.\ref{fig:2}a-d).  The different substrates  form a typical family of substrates typically known to host FM~\cite{romming2013writing,dupe2014tailoring,Simon2014,crum2015perpendicular,romming2015field,dos2016chirality,fernandes2018universality,fernandes2020defect,Arjana2020,bouhassoune2021friedel,lima2022spin}, and AFM skyrmions~\cite{aldarawsheh2022emergence,aldarawsheh2023}. The in-plane AFM Néel state is the ground state for  Mn layer in all magnetic systems, formed as a result of magnetic frustration caused by strong AFM exchange coupling among the first nearest neighbors, as illustrated in Supplementary Fig. 1. While the in-plane orientation is dictated  by the $z$ component of the antisymmetric exchange interactions (Dzyaloshinskii-Moriya Interactions -- DMI) and is further reinforced by the in-plane magnetic anisotropy energy (MAE) $K$.

\section*{Results}

\subsection{Topological magnetic states in frustrated Mn layer.}
The  Ir(111) substrate forms a triangular lattice, on which we deposit layers of Mn, PdMn, MnPdFe and MnPd$_2$Fe and perform atomistic spin dynamics~\cite{muller2019spirit}, minimizing the Heisenberg Hamiltonian (Eq.~\ref{eq.Heisenberg} in Methods section) equipped with the magnetic interactions  derived from first-principles (see Methods section).  We identify a  plethora of  AFM N\'eel  meronic magnetic states forming metastable states  in the Mn layer as depicted in Fig.~\ref{fig:1}a,  Fig.~\ref{fig:2}g, h, and Supplementary Fig. 2.

The N\'eel ordering of the spins is the ground state of the Mn layer in all of the  aforementioned magnetic systems. The associated critical temperatures range from 130K for PdMn bilayers to about 600K or more for the rest of explored Mn-based films. The spins forming the AFM N\'eel order   are segmented into  three  sublattices L1, L2 and L3, each hosting FM spin alignment  (Fig.~\ref{fig:1}j). At each sublattice, FM meronic pair can be stabilized, so in total, in the case of single AFM N\'eel meronic pair (Fig.~\ref{fig:1}a), we have six FM merons (antimerons), as shown in Fig.~\ref{fig:1}d-i, which we refer to as a hexameronic state. 
By zooming in into the two spin-swirling extremities of the hexameron (Figs.~\ref{fig:1}b-c) and their respective sublattice decomposition (Figs.~\ref{fig:1}d-i), we identify a vortex (Fig.~\ref{fig:1}d) and an antivortex (Fig.~\ref{fig:1}h) whose cores reside on an Mn lattice site, around which the spins of the remaining meronic textures precess, as dictated by the magnetic frustration induced by the underlying AFM magnetic interactions.

Each of the FM building blocks of our AFM explored solitons holds a topological charge (N) defined as: N = (wp/2)~\cite{tretiakov2007vortices}, where  w = +1 (-1) for the vortex (antivortex) is the   winding number describing the rotational direction of the in-plane magnetization, and  p is the polarity which defines the out-of-plane magnetization of the center being +1 (-1) when pointing up (down)~\cite{shinjo2000magnetic}.
Since the merons and antimerons carry a topological charge of -1/2 and +1/2, respectively~\cite{yu2018transformation,gao2019creation, gao2020fractional}, the sublattice charge N$_{L}$ is either -1 (+1) for a meron-meron (antimeron-antimeron) pair, as the case of L3 (Fig.~\ref{fig:1} f, i)}, or 0 for a hybrid (see L1 and L2 in Fig.~\ref{fig:1} d, e, g, h) meron-antimeron pair. By summing up the total charge N$_{t}$ for a hexameron, one can end up with three possible values -1, 0 and +1 (see Fig.~\ref{fig:1}k),  which interestingly are energetically degenerate in the absence of an external magnetic field.

Besides the hexameronic frustrated AFM N\'eel state, we identified a rich set of other meronic textures, such as the dodecameron, hosting 12 merons, shown in Fig.~\ref{fig:2}g. Further examples of complex multimerons are presented in Supplementary Fig. 2. Similarly to the purely FM counterparts, in confined geometries  (See Supplementary Fig. ~2b-c) a "single" AFM N\'eel meronic state can be stabilized. This object is a trimeron resulting from three frustrated merons with overlapping cores,  carrying in total a half integer topological charge.

\subsection{Stability against external magnetic fields.} The investigation of the  response of topologically paired AFM N\'eel meronic pairs to an magnetic fields is important to inspect stability aspects and to fingerprint subsequent potential non-trivial topological transitions.

The frustrated meronic textures survive to extremely high in-plane magnetic fields ($>200$ Tesla). The case of an out-of-plane (OOP) magnetic  field shows a rather rich impact on the explored spin-textures. Therefore, here, we scrutinize in detail the latter scenario by focusing on three different AFM N\'eel meronic states  (see Figs.~\ref{fig:2} f-h).

As a prototypical chiral magnetic object, we consider the hexameron emerging either in the AFM N\'eel (Fig.~\ref{fig:2}f) or in the spiraling AFM N\'eel states (Fig.~\ref{fig:2}h) as well as the dodecameron (Fig.~\ref{fig:2}g). For interfaces hosting the Fe layer, MnPdFe/Ir(111) and MnPd$_2$Fe/Ir(111), we examined  both cases: switching-off (solid bars in Fig.~\ref{fig:2}e) and -on (dashed bars in Fig.~\ref{fig:2}e) the Mn-Fe magnetic interactions. A snapshot for the Mn-hexameron interfaced with ferromagnetic Fe spirals and skyrmkion is illustrated in Fig.~\ref{fig:2}i.

While we were expecting the robustness of the unveiled meronic textures against external magnetic fields, we were intrigued by the annihilation of some hexamerons emerging in an AFM N\'eel background with experimentally accessible  OOP fields,  e.g. 10 Tesla, in contrast  to dodecamerons and hexamerons arising in a N\'eel spiraling state (red and green bars  in Fig.~\ref{fig:2}e).% \samir{why are the latter more stable?}\amal{I noticed that when the hexamerons are with N\'eel spirals at the background they survive the topological transition without collapsing as well as the presence of other merons in the cell,  so when the transition occurs, and if the state is not stable and the AFM vortex-antivortex pair starts to roll toward each other in order to annihilate, the presence of spirals or other merons will prevent that from happening, which helps the AFM structure which experienced  the topological transition to rearrange their spins   without collapsing, so  the presence of the spiralisation at the back ground or other pairs of merons provides protection to the structures.}  \samir{you explain it nicely. THis should be added to the text}

To get insight into the origin of the sensitivity of these magnetic states,hexamerons forming in an AFM N\'eel background, we scrutinize the sublattices topological distribution along with the spin orientation at each sublattice of the different hexameronic states shown in Fig.~\ref{fig:3} (See supplementary Fig. 3 illustrating snapshots of the different hexamerons). As introduced earlier, there is a quadruple degeneracy for each hexameron in the absence of a magnetic field. The four states, denoted Hexa A--D and illustrated in  Fig.~\ref{fig:3}, can be distinguished by the vortex nature of their core constituents and the orientation of the core spins (see Fig.~\ref{fig:1}k). A finite OOP field lifts partially the degeneracy and favors the hexameron, here Hexa D, with most spins pointing along the field direction (see also Supplementary Fig. 3). Among the four hexamerons, Hexa D will be the most robust to the applied field and therefore survives gigantic fields. The remaining hexamerons experience at some point magnetization switching to reach the optimal sublattice topological distribution  defined by Hexa D. This requires a flip of the spins for at least one meron (antimeron) implying going through a topological charge transition, being a non trivial process, during which, the AFM meronic structure might encounter an unstable spin distribution, leading to the annihilation of the AFM meronic structure where the AFM vortex and antivortex start rolling towards each other and then collapse at a rather low magnetic field. If the transition occurs, however, the new magnetic state would be capable of surviving large magnetic fields similar to Hexa D.

 However, the presence of N\'eel spirals in the background or additional pairs of AFM meronic textures (leading to for example to dodecamerons) prevent the formation of unstable states within the topological transition induced by the magnetic field, which would lead to the collapse of the frustrated soliton. Effectively, a barrier is provided by enabling the rearrangement of the spins to acquire the desired topological state, which would withstand immense magnetic fields.

%This prevention mechanism, facilitated by the spirals or other merons, enables the AFM structure to rearrange its spins without collapsing This resilience can be attributed to the existence of spirals or other AFM pair of merons at the cell which  prevents vortex and antivortex spins from rolling towards each other when the unstable state resulting from the transition occurs. This prevention, facilitated by the presence of spirals or other merons, aids the AFM structure in rearranging its spins without collapsing and hence survive gigantic magnetic field. }

%While for the case of the presence of N\'eel spirals at the background or more pairs of AFM meronic structures in the cell provides protection to the meronic structure and allows the topological transition go smoothly by preventing the vortex-antivortex to roll toward each other during the process which guarantees the smooth topological transition and hence being stable against the magnetic field }

\subsection{Emergence mechanism.}

We have identified that the formation of our frustrated AFM N\'eel meronic spin textures requires a strong AFM exchange coupling among the first nearest-neighbor atoms $J_{1}$ (see Supplementary Fig. 1a-d). This coupling is responsible for the AFM N\'eel order of the spins, and it is through magnetic frustration that these solitons may arise. Additionally,  another magnetic interaction is required to align the spins in the in-plane direction. This interaction can be provided by the in-plane MAE, $K$  $<$ 0, as observed in Mn/Ir(111), while for the other three magnetic systems studied, $K$ prefers an out-of-plane orientation of spins (Supplementary Fig.1e). However, the $z$ component of the DMI vector ($D_{z}$) plays a crucial role in aligning the spins in-plane, ultimately leading to the emergence of the AFM N\'eel meronic textures. In conclusion, to obtain our AFM solitons on a triangular lattice, an AFM $J_{1}$ is required, along with either a finite $D_{z}$ or an in-plane $K$.

To explore the fundamental mechanisms defining the stability of the spin-textures, we built a minimal spin model that contains only an AFM $J_1$ and $D_z$ since the latter played the main role in stabilizing the meronic textures in the four investigated Mn-based interfaces. The resulting phase diagram is shown in Fig.~\ref{fig:4}a. While the ground state would have been a pure N\'eel state without $D_z$, the latter enables quickly the formation of frustrated merons. Increasing $D_z$ enforces a stronger in-plane alignment of the spins, which reduces the size of the meronic constituents  (Fig.~\ref{fig:4}b and Supplementary Fig. 4). Clearly, the size of merons is dictated by a competition of magnetic exchange and DMI. Keeping $D_z$ fixed while increasing the AFM $J_1$ counteracts the effect of DMI and enlarges the meron core (Fig.~\ref{fig:4}c).

Fig.~\ref{fig:4}d presents the critical OOP magnetic field upon which the meronic texture, here Hexa D similar to that shown in Fig.~\ref{fig:3}, is annihilated as function of the OOP DMI component all normalized by the nearest neighboring AFM exchange interaction. The obtained curve follows a quadratic dependence, highlighting that the DMI enhances the stability of the frustrated merons.  In fact, the application of an  OOP magnetic field counteracts the influence of the OOP DMI component by tilting the spins to the OOP direction,  causing disruption to the in-plane alignment of the spins, imposed by the OOP DMI component, throughout the surrounding area, including the region spanning between the extremities of the hexameron, ultimately leading to its collapse. Consequently, the larger the OOP DMI component (smaller meronic cores), the larger the critical field required to destroy the AFM spin-swirling textures.

%\amal{In fact,  applying a OOP magnetic field counter acts the effect of OOP DMI component by tilting the spins to OOP direction, leading to disturbance for the in-plane alignment of the spins in the  background including the region between the extremities of the hexameron,
%ultimately leading to its collapse, which is countered by effect of the OOP DMI component. Consequently, the larger the OOP DMI (smaller meronic states)}
\section*{Discussion}

Our ab-initio simulations uncovered non-synthetic N\'eel-frustrated AFM meronic textures emerging in a realistic set of materials and interfaces. The newly unveiled magnetic objects are hosted by  triangular Mn layer interfaced with an Ir(111) surface, with either a Pd overlayer, separated from Ir with PdFe bilayer of Pd$_2$Fe trilayer, which all represent substrates that can readily be grown experimentally. The frustrated AFM states form hexamerons, composed of   three FM meronic pairs each located at one of the three FM sublattices building up the AFM N\'eel background. Other solitons can emerge such as dodecamerons (12 merons) while confined geometries enables the stabilization of a frustrated trimeron.

We have observed that these AFM N\'eel meronic solitons survive high values of magnetic fields if the majority spins align in the direction of the OOP magnetic field. Otherwise, a transition of the sublattice topological charge  occurs, leading to the potential annihilation of the AFM solitons at experimentally accessible values of magnetic fields. To gain a better understanding of the characteristics of these AFM solitons, we provided a spin model that outlines the minimum set of magnetic interactions necessary to generate the detected AFM solitons.

Identifying new AFM solitons with a realistic existence scenario is at the heart of AFM topological magnetism. Our predictions can initiate the experimental discovery of the intriguing  intrinsic frustrated multimeronic textures, which can delineate in various topological sequences.  It remains to be explored how such spin states can be implemented and designed in AFM spintronic devices. Certainly, the thin films being proposed provide a solid platform for AFM meronic textures with a potential impact in information technology. 

\begin{figure}[ht]
	\begin{center}
	\includegraphics[width=1\columnwidth,keepaspectratio]{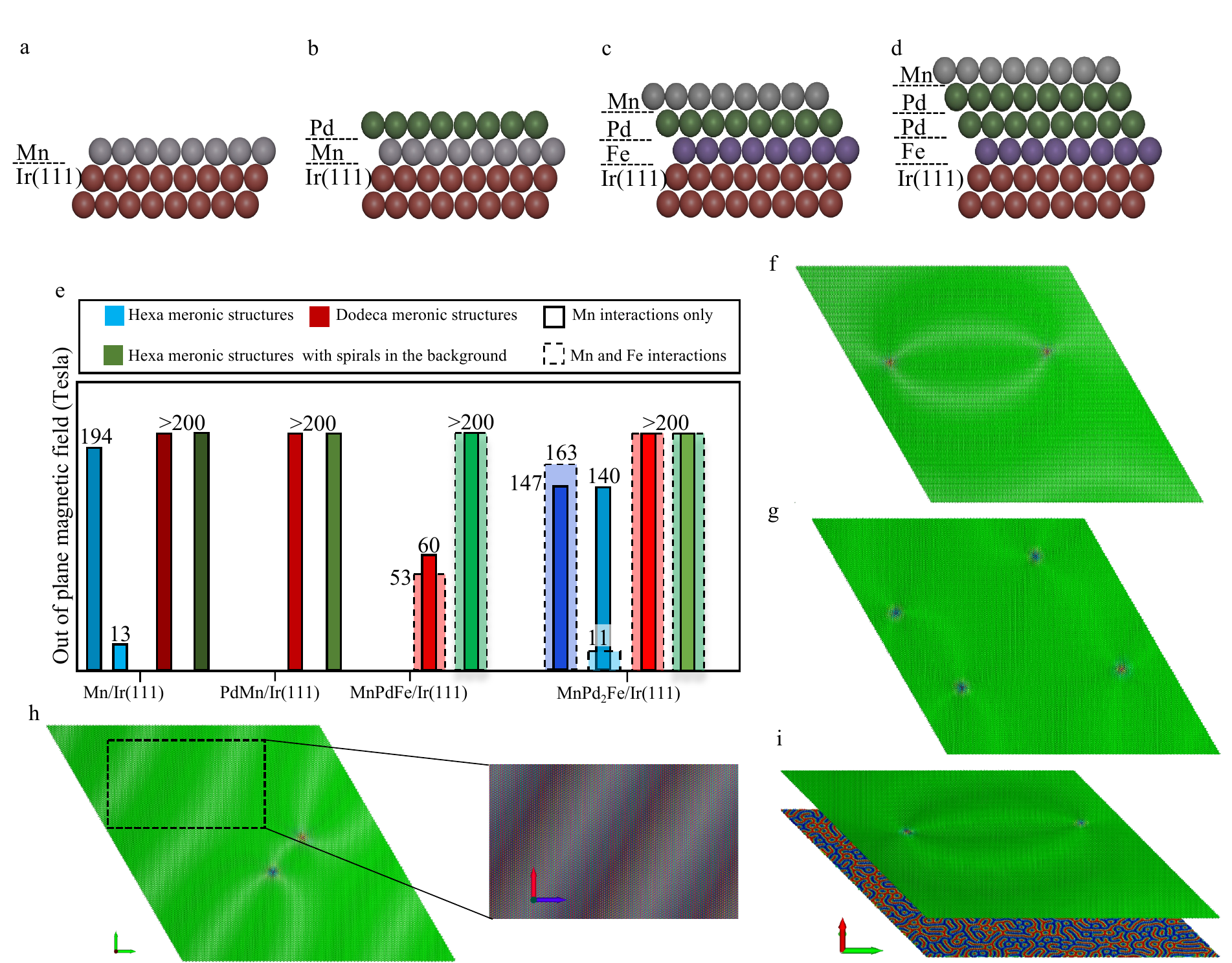}
	\caption{\textbf{Impact of external magnetic field }: \textbf{a-d} Schematic representation of the magnetic layered systems of study. \textbf{e} The critical  out-of-plane magnetic field annihilating. Blue and red bars refer to respectively hexa- and dodecamerons emerging in AFM N\'eel order, while green bars correspond to hexamerons hosted by initially spiraling N\'eel states. Bars with solid versus dashed lines distinguish the cases without and with Mn-Fe interactions. \textbf{f-h} snapshots for the blue, red and green magnetic states, respectively, inset shows the spiral at the background of \textbf{h}. \textbf{i} Snapshot showing the blue dashed bar representing the Mn layer(upper layer) interfaced with the Fe layer (lower layer).  }
		\label{fig:2}
	\end{center}
\end{figure}

\begin{methods}

In this study, we conducted a systematic investigation to explore the magnetic structures that can be hosted by  the magnetic layers of our four layered systems. Our approach involved a three-fold procedure, combining ab-initio calculations with spin atomistic dynamics. The details of this procedure are outlined below. 
\subsection{Ab-initio calculations.}

To simulate the magnetic properties of our magnetic layers, we utilized in a first step the Quantum-Espresso computational package~\cite{giannozzi2009quantum}. The calculations employed projector augmented wave pseudo potentials sourced from the PS Library~\cite{dal2014pseudopotentials} and the self-consistent calculations were performed with
a k-mesh of  $28\times28\times1$ points for the unit cell. The layers were arranged in an fcc-stacked configuration along the [111] direction (See to Fig.~\ref{fig:2}a-d). The relaxation parameters were then extracted, revealing the relaxation percentages of the different layers in relation to the ideal interlayer distance in the Ir-based systems. Specifically, for \textbf{Mn/Ir(111)}, the relaxation percentages were 2.3\% and -3.4\%; for \textbf{PdMn/Ir(111)}, they were 8.6\%, 10.3\%, and -2.3\%; for \textbf{MnPdFe/Ir(111)}, the percentages were 4\%, 5.2\%, 8.1\%, and -1\%; and for \textbf{MnPd$_2$Fe/Ir(111)}, they were 5.9\%, -4\%, 8.2\%, 8.2\%, and -0.7\%, for each layer respectively. Here, positive (negative) values indicate atomic relaxations towards (away from) the Ir surface.

After establishing the geometries of the various magnetic systems, we conducted in a second step a detailed investigation of their magnetic properties and interactions using the all-electron full-potential relativistic Korringa-Kohn-Rostoker  (KKR) Green function method, implemented in the JuKKR computational package~\cite{Papanikolaou2002,Bauer2014},in the local spin density approximation. Each of the four magnetic systems consists of a slab with 30 layers. In the case of \textbf{Mn/Ir(111)}, the slab consists of 5 vacuum + 1 Mn + 20 Ir layers + 4 vacuum layers. For \textbf{PdMn/Ir(111)}, the slab comprises 4 vacuum  + 1 Pd + 1 Mn layer+ 20 Ir layers + 4 vacuum . In the case of \textbf{MnPdFe/Ir(111)}, the slab includes 3 vacuum + 1 Mn layer + 1 Pd  + 1 Fe + 20 Ir + 4 vacuum layers. Lastly, for \textbf{MnPd$_2$Fe/Ir(111)}, the slab is composed of 2 vacuum + 1 Mn + 2 Pd + 1 Fe + 20 Ir + 4 vacuum. To perform the calculations, the momentum expansion of the Green function was truncated at $\ell_{\text{max}} = 3$. Self-consistent calculations were conducted using a k-mesh of $30\times30\times1$ points. The energy contour consisted of 23 complex energy points in the upper complex plane, and it incorporated 9 Matsubara poles.
To extract the Heisenberg exchange interactions and Dzyaloshinskii-Moriya (DM) vectors, we employed the infinitesimal rotation method ~\cite{ Liechtenstein1987,Ebert2009}. For this extraction, we used a finer k-mesh of $200\times200\times1$ points.

  \subsection{Hamiltonian Model and atomistic spin dynamics.}
%  \samir{careful with your text, needs to be different from Nature Comm}
 % After extracting the magnetic parameters for our magnetic atoms, we explore the magnetic states that can emerge at the magnetic layers utilizing spin atomistic dynamics to minimize the two dimensional  Heisenberg Hamiltonian on a triangular latttice equipped with, equipped with Heisenberg exchange coupling, DMI, the magnetic anisotropy energy, and Zeeman term. The energy functional  reads as follows:
 %Once the magnetic parameters for our magnetic atoms have been extracted, to explore the magnetic properties and emerging complex states we utilize the Landau-Lifshitz-equation (LLG) as implemented in the Spirit code~\cite{muller2019spirit}. This is accomplished by minimizing the two-dimensional Heisenberg Hamiltonian on a triangular lattice. The Hamiltonian incorporates various terms, including Heisenberg exchange coupling, Dzyaloshinskii-Moriya interaction (DMI), magnetic anisotropy energy, and Zeeman term. The energy functional of the system can be described as follows:
After extracting the magnetic parameters for our magnetic atoms from first-principles, we employ the Landau-Lifshitz equation (LLG) implemented in the Spirit code~\cite{muller2019spirit} to explore the magnetic properties and complex states. This exploration involves minimizing the two-dimensional Heisenberg Hamiltonian on a triangular lattice. The Hamiltonian comprises several terms, including Heisenberg exchange coupling, Dzyaloshinskii-Moriya interaction (DMI), magnetic anisotropy energy, and Zeeman term. The energy functional of the system can be described as follows:
 \begin{equation}
   H = H_\text{Exchange} + H_\text{DMI} + H_\text{Anisotropy} + H_\text{Zeeman},
   \label{eq.Heisenberg}
\end{equation}
with
\begin{equation*}
 H_\text{Exchange}= -\sum\limits_{<i,j>} J^\text{Mn-Mn}_{ij}\; \textbf{S}_{i}\cdot \textbf{S}_{j}       -\sum\limits_{<i,j>} J^\text{Fe-Mn}_{ij}\; \textbf{S}_{i}\cdot \textbf{S}_{j}   -\sum\limits_{<i,j>} J^\text{Fe-Fe}_{ij}\;\textbf{S}_{i}\cdot \textbf{S}_{j},
\end{equation*}
\begin{equation*}
 H_\text{DMI}= -\sum\limits_{<i,j>}\textbf{D}^\text{Mn-Mn}_{ij}\cdot [\textbf{S}_{i}\times \textbf{S}_{j}]-\sum\limits_{<i,j>}\textbf{D}^\text{Fe-Mn}_{ij}\cdot [\textbf{S}_{i}\times \textbf{S}_{j}]-\sum\limits_{<i,j>}\textbf{D}^\text{Fe-Fe}_{ij}\cdot [\textbf{S}_{i}\times \textbf{S}_{j}],  
 \end{equation*}
 \begin{equation*}
 H_\text{Anisotropy}=- K^\text{Mn}\sum\limits_{i}  (\textbf{S}_{i} \cdot \textbf{e}_{i})^2 - K^\text{Fe}\sum\limits_{i}  (\textbf{S}_{i} \cdot \textbf{e}_{i})^2,
 \end{equation*}
 \begin{equation*}
H_\text{Zeeman} =- \sum\limits_{i} \mu_i\textbf{B}\cdot \textbf{S}_i,  
\end{equation*}
where  we assign indices $i$ and $j$ to denote specific sites, each associated with a magnetic moment. The magnetic moment is represented by the unit vector $\textbf{S}$. The Heisenberg exchange coupling strength $J^\text{X-Y}_{ij}$ describes the interaction between an atom X on site $i$ and an atom Y on site $j$, where a negative value indicates AFM interaction. Similarly, we use the notation $\textbf{ D}$ for the Dzyaloshinskii-Moriya interaction vector, $ K$ for the magnetic anisotropy energy, and $\mu_i\textbf{B}$ to represent the Zeeman coupling to the atomic spin moment $\mu$ at site $i$. It is important to note that the Fe-Mn and Fe-Fe interactions are only considered in the \textbf{MnPdFe/Ir(111)} and \textbf{MnPd$_2$Fe/Ir(111)} systems. For our spin atomistic simulations, we adopt both periodic and finite boundary conditions to model the extended and confined two-dimensional system, respectively, with cells containing  $249^2$, $300^2$, $390^2$ sites.

  %$i$ and $j$ are site indices carrying each magnetic moments.  $\textbf{S}$ is a unit vector of the magnetic moment. $J^\text{X-Y}_{ij}$ is the Heisenberg exchange coupling strength, being $<$ 0 for AFM interaction, between an X atom on site $i$ and a Y atom on site $j$. A similar notation is adopted for the DMI vector $\textbf{D}$ and the magnetic anisotropy energy $K$ and $ \mu_i \textbf{B}$  describes the Zeeman coupling to the atomic spin moment $\mu$ at site $i$. We note that obviously the Fe-Mn and Fe-Fe interactions are only included for both \textbf{MnPdFe/Ir(111)} and \textbf{MnPd$_2$Fe/Ir(111)} systems.
%In our spin atomistic simulations, we assumed periodic and finite boundary conditions  to model the extended and confined two-dimensional system with cells containing  $249^2$, $300^2$, $ 390^2$ sites. 

\subsection{Data availability}
The data needed to evaluate the conclusions in the paper are present in the paper and the Supplementary Information.

\subsection{Code availability} We used the following codes:
Quantum ESPRESSO which can be found at \url{https:/www.quantum-espresso.org/download},  SPIRIT can be found at \url{https://github.com/spirit-code/spirit}, and 
the KKR code is a rather complex ab-initio DFT-based code, which is in general impossible to use without proper training on the theory behind it and on the practical utilization of the code. We are happy to provide the latter code upon request.

\end{methods}

\begin{figure}[ht]
	\begin{center}
	\includegraphics[width=1\columnwidth,keepaspectratio]{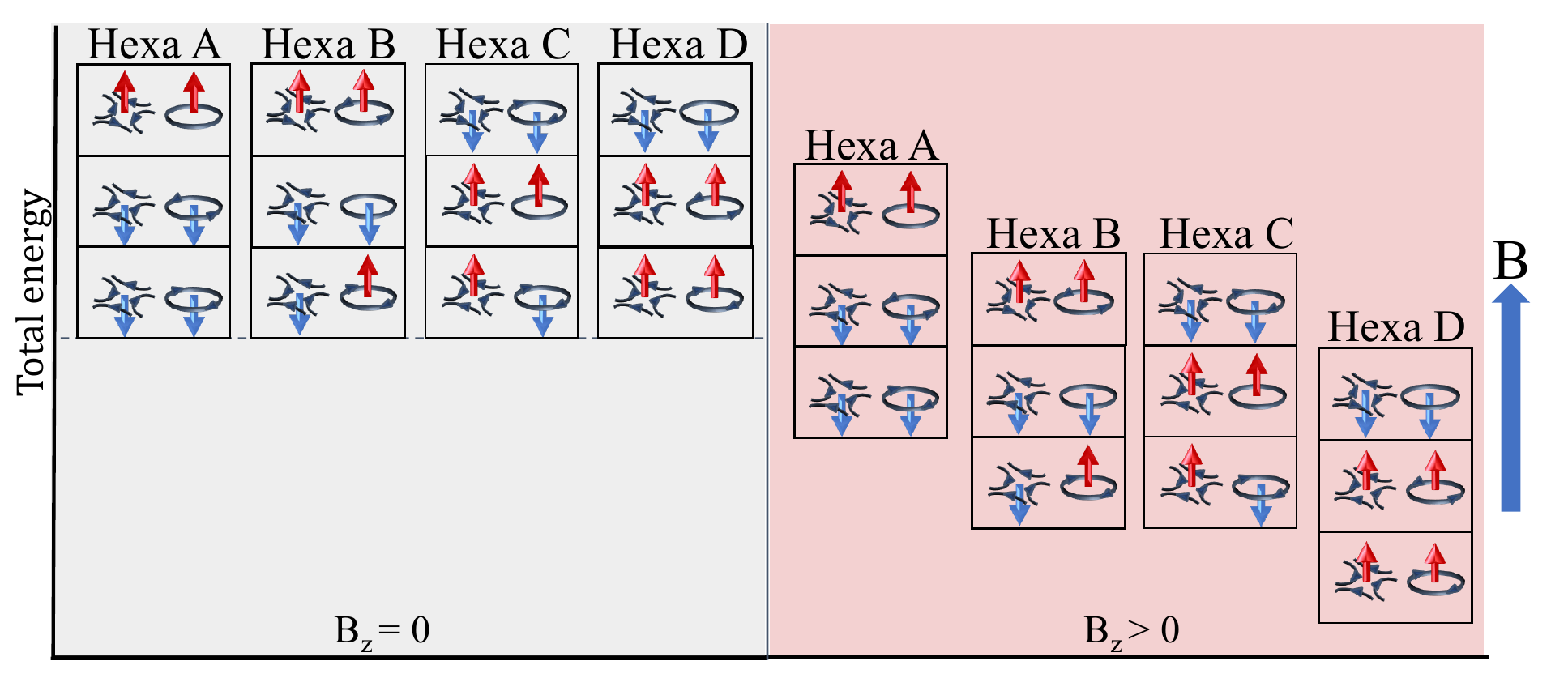} 
	\caption{\textbf{Topologically-dependent response to the external magnetic field : } 
	Lifting the quadrupole degeneracy of the hexameron (Hexa A--D) upon application of an outof-plane magnetic field. Each hexmeron is decomposed into the three sublattices with the illustration of the vortex nature of the meronic core constituents together with the core spin-direction.  Hexa D is the frustrated hexameron satisfying the ideal stability criterion against the magnetic field.   }
		\label{fig:3}
	\end{center}
\end{figure}

\begin{figure}[ht]
	\begin{center}
	\includegraphics[width=1\columnwidth,keepaspectratio]{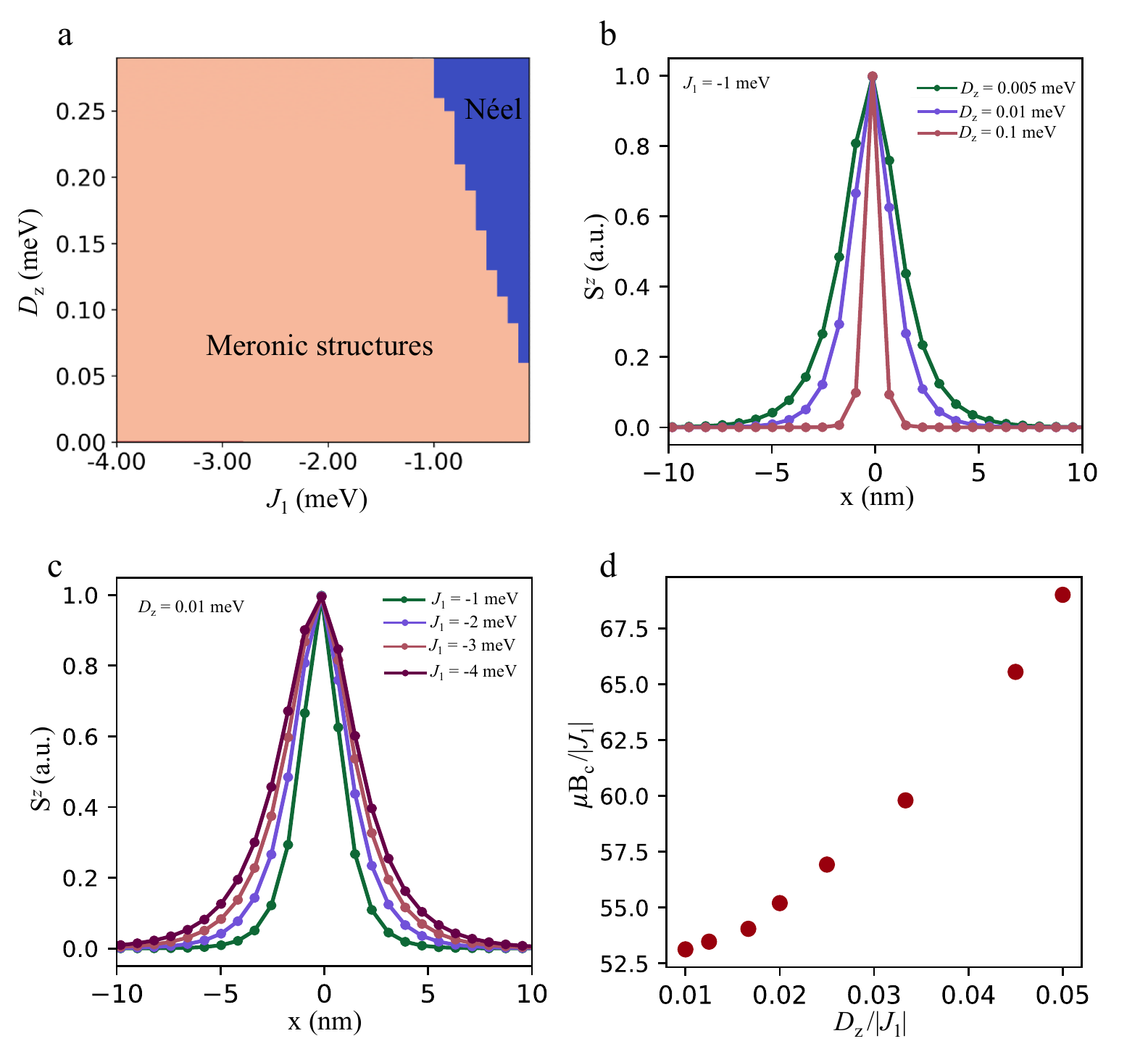} 
	\caption{\textbf{Minimal spin model for frustrated AFM N\'eel multimerons: }\textbf{a} Phase diagram  showing immediate emergence of hexamerons requiring an AFM nearest neighboring magnetic interaction, $J_1$, and an out-of-plane DMI component, $D_z$.  \textbf{b, c} Mutual impact of magnetic exchange interactions and DMI on the S${^z}$ profile of one of the vortices. \textbf{d} The associated critical field required for the annihilation of the multimeron as function of the DMI.  }
		\label{fig:4}
	\end{center}
\end{figure}
\section*{ACKNOWLEDGEMENTS}
We acknowledge fruitful discussions with Nihad Abuawwad. This work was supported by the Federal Ministry of Education and Research of Germany
in the framework of the Palestinian-German Science Bridge (BMBF grant number
01DH16027) and the Deutsche For\-schungs\-gemeinschaft (DFG) through SPP 2137 ``Skyrmionics'' (Project LO 1659/8-1). 
The authors gratefully acknowledge
the computing time granted through JARA on the supercomputer JURECA %\cite{ krause2018jureca}  
at Forschungszentrum Jülich.

%\cite{krause2018jureca}.
\begin{addendum}

\item[Author contributions]
S.L. initiated, designed and supervised the project. A.A. performed the simulations and post processed the data . A.A., M.S., M.A., and S.L. discussed the results. A.A. and S.L. wrote the manuscript to which all co-authors contributed.
\\

\item[Competing Interests]
The authors declare no competing interests.

\item[Correspondence] Correspondence and requests for materials should be addressed to A.A. (email: a.aldarawsheh@fz-juelich.de) or to S.L. (email: s.lounis@fz-juelich.de).

\end{addendum}


\section*{References}


\begin{thebibliography}{10}
	\expandafter\ifx\csname url\endcsname\relax
	\def\url#1{\texttt{#1}}\fi
	\expandafter\ifx\csname urlprefix\endcsname\relax\def\urlprefix{URL }\fi
	\providecommand{\bibinfo}[2]{#2}
	\providecommand{\eprint}[2][]{\url{#2}}
	
	\bibitem{Jungwirth2016}
	\bibinfo{author}{Jungwirth, T.}, \bibinfo{author}{Marti, X.},
	\bibinfo{author}{Wadley, P.} \& \bibinfo{author}{Wunderlich, J.}
	\newblock \bibinfo{title}{Antiferromagnetic spintronics}.
	\newblock \emph{\bibinfo{journal}{Nat. Nanotechnol.}}
	\textbf{\bibinfo{volume}{11}}, \bibinfo{pages}{231--241}
	(\bibinfo{year}{2016}).
	
	\bibitem{gomonay2017concepts}
	\bibinfo{author}{Gomonay, O.}, \bibinfo{author}{Jungwirth, T.} \&
	\bibinfo{author}{Sinova, J.}
	\newblock \bibinfo{title}{Concepts of antiferromagnetic spintronics}.
	\newblock \emph{\bibinfo{journal}{Phys. Stat. Sol. RRL.}}
	\textbf{\bibinfo{volume}{11}}, \bibinfo{pages}{1700022}
	(\bibinfo{year}{2017}).
	
	\bibitem{baltz2018antiferromagnetic}
	\bibinfo{author}{Baltz, V.} \emph{et~al.}
	\newblock \bibinfo{title}{Antiferromagnetic spintronics}.
	\newblock \emph{\bibinfo{journal}{Rev. Mod. Phys.}}
	\textbf{\bibinfo{volume}{90}}, \bibinfo{pages}{015005}
	(\bibinfo{year}{2018}).
	
	\bibitem{gomonay2018antiferromagnetic}
	\bibinfo{author}{Gomonay, O.}, \bibinfo{author}{Baltz, V.},
	\bibinfo{author}{Brataas, A.} \& \bibinfo{author}{Tserkovnyak, Y.}
	\newblock \bibinfo{title}{Antiferromagnetic spin textures and dynamics}.
	\newblock \emph{\bibinfo{journal}{Nat. Phys.}} \textbf{\bibinfo{volume}{14}},
	\bibinfo{pages}{213} (\bibinfo{year}{2018}).
	
	\bibitem{vsmejkal2018topological}
	\bibinfo{author}{{\v{S}}mejkal, L.}, \bibinfo{author}{Mokrousov, Y.},
	\bibinfo{author}{Yan, B.} \& \bibinfo{author}{MacDonald, A.~H.}
	\newblock \bibinfo{title}{Topological antiferromagnetic spintronics}.
	\newblock \emph{\bibinfo{journal}{Nat. Phys.}} \textbf{\bibinfo{volume}{14}},
	\bibinfo{pages}{242--251} (\bibinfo{year}{2018}).
	
	\bibitem{nvemec2018antiferromagnetic}
	\bibinfo{author}{N{\v{e}}mec, P.}, \bibinfo{author}{Fiebig, M.},
	\bibinfo{author}{Kampfrath, T.} \& \bibinfo{author}{Kimel, A.~V.}
	\newblock \bibinfo{title}{Antiferromagnetic opto-spintronics}.
	\newblock \emph{\bibinfo{journal}{Nat. Phys.}} \textbf{\bibinfo{volume}{14}},
	\bibinfo{pages}{229--241} (\bibinfo{year}{2018}).
	
	\bibitem{legrand2020room}
	\bibinfo{author}{Legrand, W.} \emph{et~al.}
	\newblock \bibinfo{title}{Room-temperature stabilization of antiferromagnetic
		skyrmions in synthetic antiferromagnets}.
	\newblock \emph{\bibinfo{journal}{Nat. Mater.}} \textbf{\bibinfo{volume}{19}},
	\bibinfo{pages}{34} (\bibinfo{year}{2020}).
	
	\bibitem{Jani2021}
	\bibinfo{author}{Jani, H.} \emph{et~al.}
	\newblock \bibinfo{title}{Antiferromagnetic half-skyrmions and bimerons at room
		temperature}.
	\newblock \emph{\bibinfo{journal}{Nature}} \textbf{\bibinfo{volume}{590}},
	\bibinfo{pages}{74} (\bibinfo{year}{2021}).
	
	\bibitem{aldarawsheh2022emergence}
	\bibinfo{author}{Aldarawsheh, A.} \emph{et~al.}
	\newblock \bibinfo{title}{Emergence of zero-field non-synthetic single and
		interchained antiferromagnetic skyrmions in thin films}.
	\newblock \emph{\bibinfo{journal}{Nat. Commun.}} \textbf{\bibinfo{volume}{13}},
	\bibinfo{pages}{7369} (\bibinfo{year}{2022}).
	
	\bibitem{dohi2019formation}
	\bibinfo{author}{Dohi, T.}, \bibinfo{author}{DuttaGupta, S.},
	\bibinfo{author}{Fukami, S.} \& \bibinfo{author}{Ohno, H.}
	\newblock \bibinfo{title}{Formation and current-induced motion of synthetic
		antiferromagnetic skyrmion bubbles}.
	\newblock \emph{\bibinfo{journal}{Nat. Commun.}} \textbf{\bibinfo{volume}{10}},
	\bibinfo{pages}{5153} (\bibinfo{year}{2019}).
	
	\bibitem{finco2021imaging}
	\bibinfo{author}{Finco, A.} \emph{et~al.}
	\newblock \bibinfo{title}{Imaging non-collinear antiferromagnetic textures via
		single spin relaxometry}.
	\newblock \emph{\bibinfo{journal}{Nat. Commun.}} \textbf{\bibinfo{volume}{12}},
	\bibinfo{pages}{767} (\bibinfo{year}{2021}).
	
	\bibitem{juge2022skyrmions}
	\bibinfo{author}{Juge, R.} \emph{et~al.}
	\newblock \bibinfo{title}{Skyrmions in synthetic antiferromagnets and their
		nucleation via electrical current and ultra-fast laser illumination}.
	\newblock \emph{\bibinfo{journal}{Nat. Commun.}} \textbf{\bibinfo{volume}{13}},
	\bibinfo{pages}{4807} (\bibinfo{year}{2022}).
	
	\bibitem{chen2022controllable}
	\bibinfo{author}{Chen, R.} \emph{et~al.}
	\newblock \bibinfo{title}{Controllable generation of antiferromagnetic
		skyrmions in synthetic antiferromagnets with thermal effect}.
	\newblock \emph{\bibinfo{journal}{Adv. Funct. Mater.}}
	\textbf{\bibinfo{volume}{32}}, \bibinfo{pages}{2111906}
	(\bibinfo{year}{2022}).
	
	\bibitem{barker2016static}
	\bibinfo{author}{Barker, J.} \& \bibinfo{author}{Tretiakov, O.~A.}
	\newblock \bibinfo{title}{Static and dynamical properties of antiferromagnetic
		skyrmions in the presence of applied current and temperature}.
	\newblock \emph{\bibinfo{journal}{Phys. Rev. Lett.}}
	\textbf{\bibinfo{volume}{116}}, \bibinfo{pages}{147203}
	(\bibinfo{year}{2016}).
	
	\bibitem{olejnik2018terahertz}
	\bibinfo{author}{Olejn{\'\i}k, K.} \emph{et~al.}
	\newblock \bibinfo{title}{Terahertz electrical writing speed in an
		antiferromagnetic memory}.
	\newblock \emph{\bibinfo{journal}{Sci. Adv.}} \textbf{\bibinfo{volume}{4}},
	\bibinfo{pages}{eaar3566} (\bibinfo{year}{2018}).
	
	\bibitem{aldarawsheh2023}
	\bibinfo{author}{Aldarawsheh, A.}, \bibinfo{author}{Sallermann, M.},
	\bibinfo{author}{Abusaa, M.} \& \bibinfo{author}{Lounis, S.}
	\newblock \bibinfo{title}{A spin model for intrinsic antiferromagnetic
		skyrmions on a triangular lattice}.
	\newblock \emph{\bibinfo{journal}{Front. Phys.}} \textbf{\bibinfo{volume}{11}}
	(\bibinfo{year}{2023}).
	
	\bibitem{kampfrath2011coherent}
	\bibinfo{author}{Kampfrath, T.} \emph{et~al.}
	\newblock \bibinfo{title}{Coherent terahertz control of antiferromagnetic spin
		waves}.
	\newblock \emph{\bibinfo{journal}{Nat. Photonics}}
	\textbf{\bibinfo{volume}{5}}, \bibinfo{pages}{31--34} (\bibinfo{year}{2011}).
	
	\bibitem{gomonay2014spintronics}
	\bibinfo{author}{Gomonay, E.} \& \bibinfo{author}{Loktev, V.}
	\newblock \bibinfo{title}{Spintronics of antiferromagnetic systems}.
	\newblock \emph{\bibinfo{journal}{Low Temp. Phys.}}
	\textbf{\bibinfo{volume}{40}}, \bibinfo{pages}{17--35}
	(\bibinfo{year}{2014}).
	
	\bibitem{baierl2016terahertz}
	\bibinfo{author}{Baierl, S.} \emph{et~al.}
	\newblock \bibinfo{title}{Terahertz-driven nonlinear spin response of
		antiferromagnetic nickel oxide}.
	\newblock \emph{\bibinfo{journal}{Phys. Rev. Lett.}}
	\textbf{\bibinfo{volume}{117}}, \bibinfo{pages}{197201}
	(\bibinfo{year}{2016}).
	
	\bibitem{bhattacharjee2018neel}
	\bibinfo{author}{Bhattacharjee, N.} \emph{et~al.}
	\newblock \bibinfo{title}{N{\'e}el spin-orbit torque driven antiferromagnetic
		resonance in mn 2 au probed by time-domain thz spectroscopy}.
	\newblock \emph{\bibinfo{journal}{Phys. Rev. Lett.}}
	\textbf{\bibinfo{volume}{120}}, \bibinfo{pages}{237201}
	(\bibinfo{year}{2018}).
	
	\bibitem{rosales2015three}
	\bibinfo{author}{Rosales, H.~D.}, \bibinfo{author}{Cabra, D.~C.} \&
	\bibinfo{author}{Pujol, P.}
	\newblock \bibinfo{title}{Three-sublattice skyrmion crystal in the
		antiferromagnetic triangular lattice}.
	\newblock \emph{\bibinfo{journal}{Phys. Rev. B}} \textbf{\bibinfo{volume}{92}},
	\bibinfo{pages}{214439} (\bibinfo{year}{2015}).
	
	\bibitem{zhang2016antiferromagnetic}
	\bibinfo{author}{Zhang, X.}, \bibinfo{author}{Zhou, Y.} \&
	\bibinfo{author}{Ezawa, M.}
	\newblock \bibinfo{title}{Antiferromagnetic skyrmion: stability, creation and
		manipulation}.
	\newblock \emph{\bibinfo{journal}{Sci. Rep.}} \textbf{\bibinfo{volume}{6}},
	\bibinfo{pages}{24795} (\bibinfo{year}{2016}).
	
	\bibitem{velkov2016phenomenology}
	\bibinfo{author}{Velkov, H.} \emph{et~al.}
	\newblock \bibinfo{title}{Phenomenology of current-induced skyrmion motion in
		antiferromagnets}.
	\newblock \emph{\bibinfo{journal}{New J. Phys.}} \textbf{\bibinfo{volume}{18}},
	\bibinfo{pages}{075016} (\bibinfo{year}{2016}).
	
	\bibitem{keesman2016skyrmions}
	\bibinfo{author}{Keesman, R.}, \bibinfo{author}{Raaijmakers, M.},
	\bibinfo{author}{Baerends, A.}, \bibinfo{author}{Barkema, G.} \&
	\bibinfo{author}{Duine, R.}
	\newblock \bibinfo{title}{Skyrmions in square-lattice antiferromagnets}.
	\newblock \emph{\bibinfo{journal}{Phys. Rev. B}} \textbf{\bibinfo{volume}{94}},
	\bibinfo{pages}{054402} (\bibinfo{year}{2016}).
	
	\bibitem{jin2016dynamics}
	\bibinfo{author}{Jin, C.}, \bibinfo{author}{Song, C.}, \bibinfo{author}{Wang,
		J.} \& \bibinfo{author}{Liu, Q.}
	\newblock \bibinfo{title}{Dynamics of antiferromagnetic skyrmion driven by the
		spin hall effect}.
	\newblock \emph{\bibinfo{journal}{Appl. Phys. Lett.}}
	\textbf{\bibinfo{volume}{109}}, \bibinfo{pages}{182404}
	(\bibinfo{year}{2016}).
	
	\bibitem{gobel2017antiferromagnetic}
	\bibinfo{author}{G{\"o}bel, B.}, \bibinfo{author}{Mook, A.},
	\bibinfo{author}{Henk, J.} \& \bibinfo{author}{Mertig, I.}
	\newblock \bibinfo{title}{Antiferromagnetic skyrmion crystals: Generation,
		topological hall, and topological spin hall effect}.
	\newblock \emph{\bibinfo{journal}{Phys. Rev. B}} \textbf{\bibinfo{volume}{96}},
	\bibinfo{pages}{060406} (\bibinfo{year}{2017}).
	
	\bibitem{tomasello2017performance}
	\bibinfo{author}{Tomasello, R.} \emph{et~al.}
	\newblock \bibinfo{title}{Performance of synthetic antiferromagnetic racetrack
		memory: domain wall versus skyrmion}.
	\newblock \emph{\bibinfo{journal}{J. Phys. D}} \textbf{\bibinfo{volume}{50}},
	\bibinfo{pages}{325302} (\bibinfo{year}{2017}).
	
	\bibitem{akosa2018theory}
	\bibinfo{author}{Akosa, C.~A.}, \bibinfo{author}{Tretiakov, O.},
	\bibinfo{author}{Tatara, G.} \& \bibinfo{author}{Manchon, A.}
	\newblock \bibinfo{title}{Theory of the topological spin hall effect in
		antiferromagnetic skyrmions: Impact on current-induced motion}.
	\newblock \emph{\bibinfo{journal}{Phys. Rev. Lett.}}
	\textbf{\bibinfo{volume}{121}}, \bibinfo{pages}{097204}
	(\bibinfo{year}{2018}).
	
	\bibitem{silva2019antiferromagnetic}
	\bibinfo{author}{Silva, R.}, \bibinfo{author}{Silva, R.},
	\bibinfo{author}{Pereira, A.} \& \bibinfo{author}{Moura-Melo, W.}
	\newblock \bibinfo{title}{Antiferromagnetic skyrmions overcoming obstacles in a
		racetrack}.
	\newblock \emph{\bibinfo{journal}{J. Phys.: Condens. Matter}}
	\textbf{\bibinfo{volume}{31}}, \bibinfo{pages}{225802}
	(\bibinfo{year}{2019}).
	
	\bibitem{fernandes2019skyrmions}
	\bibinfo{author}{Fernandes, R.}, \bibinfo{author}{Lopes, R.} \&
	\bibinfo{author}{Pereira, A.}
	\newblock \bibinfo{title}{Skyrmions and merons in two-dimensional
		antiferromagnetic systems}.
	\newblock \emph{\bibinfo{journal}{Solid State Commun.}}
	\textbf{\bibinfo{volume}{290}}, \bibinfo{pages}{55--59}
	(\bibinfo{year}{2019}).
	
	\bibitem{gao2020fractional}
	\bibinfo{author}{Gao, S.} \emph{et~al.}
	\newblock \bibinfo{title}{Fractional antiferromagnetic skyrmion lattice induced
		by anisotropic couplings}.
	\newblock \emph{\bibinfo{journal}{Nature}} \textbf{\bibinfo{volume}{586}},
	\bibinfo{pages}{37} (\bibinfo{year}{2020}).
	
	\bibitem{li2020bimeron}
	\bibinfo{author}{Li, X.} \emph{et~al.}
	\newblock \bibinfo{title}{Bimeron clusters in chiral antiferromagnets}.
	\newblock \emph{\bibinfo{journal}{npj Comp. Mater.}}
	\textbf{\bibinfo{volume}{6}}, \bibinfo{pages}{169} (\bibinfo{year}{2020}).
	
	\bibitem{shen2020current}
	\bibinfo{author}{Shen, L.} \emph{et~al.}
	\newblock \bibinfo{title}{Current-induced dynamics and chaos of
		antiferromagnetic bimerons}.
	\newblock \emph{\bibinfo{journal}{Phys. Rev. Lett.}}
	\textbf{\bibinfo{volume}{124}}, \bibinfo{pages}{037202}
	(\bibinfo{year}{2020}).
	
	\bibitem{silva2021antiferromagnetic}
	\bibinfo{author}{Silva, R.}
	\newblock \bibinfo{title}{Antiferromagnetic-bimeron dynamics driven by a
		spin-polarized current at an inhomogeneous racetrack}.
	\newblock \emph{\bibinfo{journal}{Phys. Lett. A}}
	\textbf{\bibinfo{volume}{403}}, \bibinfo{pages}{127399}
	(\bibinfo{year}{2021}).
	
	\bibitem{Lin2013}
	\bibinfo{author}{Lin, S.-Z.}, \bibinfo{author}{Reichhardt, C.},
	\bibinfo{author}{Batista, C.~D.} \& \bibinfo{author}{Saxena, A.}
	\newblock \bibinfo{title}{Particle model for skyrmions in metallic chiral
		magnets: Dynamics, pinning, and creep}.
	\newblock \emph{\bibinfo{journal}{Phys. Rev. B}} \textbf{\bibinfo{volume}{87}},
	\bibinfo{pages}{214419} (\bibinfo{year}{2013}).
	
	\bibitem{Nagaosa2013}
	\bibinfo{author}{Nagaosa, N.} \& \bibinfo{author}{Tokura, Y.}
	\newblock \bibinfo{title}{{Topological properties and dynamics of magnetic
			skyrmions.}}
	\newblock \emph{\bibinfo{journal}{Nat. Nanotechnol.}}
	\textbf{\bibinfo{volume}{8}}, \bibinfo{pages}{899} (\bibinfo{year}{2013}).
	
	\bibitem{Jiang2016}
	\bibinfo{author}{Jiang, W.} \emph{et~al.}
	\newblock \bibinfo{title}{{Direct observation of the skyrmion Hall effect}}.
	\newblock \emph{\bibinfo{journal}{Nat. Phys.}} \textbf{\bibinfo{volume}{13}},
	\bibinfo{pages}{162} (\bibinfo{year}{2017}).
	
	\bibitem{woo2016observation}
	\bibinfo{author}{Woo, S.} \emph{et~al.}
	\newblock \bibinfo{title}{Observation of room-temperature magnetic skyrmions
		and their current-driven dynamics in ultrathin metallic ferromagnets}.
	\newblock \emph{\bibinfo{journal}{Nat. Mater.}} \textbf{\bibinfo{volume}{15}},
	\bibinfo{pages}{501--506} (\bibinfo{year}{2016}).
	
	\bibitem{Litzius2017}
	\bibinfo{author}{Litzius, K.} \emph{et~al.}
	\newblock \bibinfo{title}{Skyrmion hall effect revealed by direct time-resolved
		x-ray microscopy}.
	\newblock \emph{\bibinfo{journal}{Nat. Phys.}} \textbf{\bibinfo{volume}{13}},
	\bibinfo{pages}{170--175} (\bibinfo{year}{2017}).
	
	\bibitem{tretiakov2007vortices}
	\bibinfo{author}{Tretiakov, O.} \& \bibinfo{author}{Tchernyshyov, O.}
	\newblock \bibinfo{title}{Vortices in thin ferromagnetic films and the skyrmion
		number}.
	\newblock \emph{\bibinfo{journal}{Phys. Rev. B}} \textbf{\bibinfo{volume}{75}},
	\bibinfo{pages}{012408} (\bibinfo{year}{2007}).
	
	\bibitem{ezawa2011compact}
	\bibinfo{author}{Ezawa, M.}
	\newblock \bibinfo{title}{Compact merons and skyrmions in thin chiral magnetic
		films}.
	\newblock \emph{\bibinfo{journal}{Phys. Rev. B}} \textbf{\bibinfo{volume}{83}},
	\bibinfo{pages}{100408} (\bibinfo{year}{2011}).
	
	\bibitem{phatak2012direct}
	\bibinfo{author}{Phatak, C.}, \bibinfo{author}{Petford-Long, A.} \&
	\bibinfo{author}{Heinonen, O.}
	\newblock \bibinfo{title}{Direct observation of unconventional topological spin
		structure in coupled magnetic discs}.
	\newblock \emph{\bibinfo{journal}{Phys. Rev. Lett.}}
	\textbf{\bibinfo{volume}{108}}, \bibinfo{pages}{067205}
	(\bibinfo{year}{2012}).
	
	\bibitem{lin2015skyrmion}
	\bibinfo{author}{Lin, S.-Z.}, \bibinfo{author}{Saxena, A.} \&
	\bibinfo{author}{Batista, C.~D.}
	\newblock \bibinfo{title}{Skyrmion fractionalization and merons in chiral
		magnets with easy-plane anisotropy}.
	\newblock \emph{\bibinfo{journal}{Phys. Rev. B}} \textbf{\bibinfo{volume}{91}},
	\bibinfo{pages}{224407} (\bibinfo{year}{2015}).
	
	\bibitem{tan2016topology}
	\bibinfo{author}{Tan, A.} \emph{et~al.}
	\newblock \bibinfo{title}{Topology of spin meron pairs in coupled {Ni/Fe/Co/Cu}
		(001) disks}.
	\newblock \emph{\bibinfo{journal}{Phys. Rev. B}} \textbf{\bibinfo{volume}{94}},
	\bibinfo{pages}{014433} (\bibinfo{year}{2016}).
	
	\bibitem{yu2018transformation}
	\bibinfo{author}{Yu, X.} \emph{et~al.}
	\newblock \bibinfo{title}{Transformation between meron and skyrmion topological
		spin textures in a chiral magnet}.
	\newblock \emph{\bibinfo{journal}{Nature}} \textbf{\bibinfo{volume}{564}},
	\bibinfo{pages}{95--98} (\bibinfo{year}{2018}).
	
	\bibitem{lu2020meron}
	\bibinfo{author}{Lu, X.}, \bibinfo{author}{Fei, R.}, \bibinfo{author}{Zhu, L.}
	\& \bibinfo{author}{Yang, L.}
	\newblock \bibinfo{title}{Meron-like topological spin defects in monolayer
		${C}r{C}l_3$}.
	\newblock \emph{\bibinfo{journal}{Nat. Commun.}} \textbf{\bibinfo{volume}{11}},
	\bibinfo{pages}{4724} (\bibinfo{year}{2020}).
	
	\bibitem{augustin2021properties}
	\bibinfo{author}{Augustin, M.}, \bibinfo{author}{Jenkins, S.},
	\bibinfo{author}{Evans, R.~F.}, \bibinfo{author}{Novoselov, K.~S.} \&
	\bibinfo{author}{Santos, E.~J.}
	\newblock \bibinfo{title}{Properties and dynamics of meron topological spin
		textures in the two-dimensional magnet ${C}r{C}l_{3}$}.
	\newblock \emph{\bibinfo{journal}{Nat. Commun.}} \textbf{\bibinfo{volume}{12}},
	\bibinfo{pages}{185} (\bibinfo{year}{2021}).
	
	\bibitem{hayami2021meron}
	\bibinfo{author}{Hayami, S.} \& \bibinfo{author}{Yambe, R.}
	\newblock \bibinfo{title}{Meron-antimeron crystals in noncentrosymmetric
		itinerant magnets on a triangular lattice}.
	\newblock \emph{\bibinfo{journal}{Phys. Rev. B}}
	\textbf{\bibinfo{volume}{104}}, \bibinfo{pages}{094425}
	(\bibinfo{year}{2021}).
	
	\bibitem{xia2022qubits}
	\bibinfo{author}{Xia, J.}, \bibinfo{author}{Zhang, X.}, \bibinfo{author}{Liu,
		X.}, \bibinfo{author}{Zhou, Y.} \& \bibinfo{author}{Ezawa, M.}
	\newblock \bibinfo{title}{Qubits based on merons in magnetic nanodisks}.
	\newblock \emph{\bibinfo{journal}{Commun. Mater.}}
	\textbf{\bibinfo{volume}{3}}, \bibinfo{pages}{88} (\bibinfo{year}{2022}).
	
	\bibitem{donnelly2021experimental}
	\bibinfo{author}{Donnelly, C.} \emph{et~al.}
	\newblock \bibinfo{title}{Experimental observation of vortex rings in a bulk
		magnet}.
	\newblock \emph{\bibinfo{journal}{Nat. Phys.}} \textbf{\bibinfo{volume}{17}},
	\bibinfo{pages}{316--321} (\bibinfo{year}{2021}).
	
	\bibitem{gao2019creation}
	\bibinfo{author}{Gao, N.} \emph{et~al.}
	\newblock \bibinfo{title}{Creation and annihilation of topological meron pairs
		in in-plane magnetized films}.
	\newblock \emph{\bibinfo{journal}{Nat. Commun.}} \textbf{\bibinfo{volume}{10}},
	\bibinfo{pages}{5603} (\bibinfo{year}{2019}).
	
	\bibitem{kolesnikov2018composite}
	\bibinfo{author}{Kolesnikov, A.} \emph{et~al.}
	\newblock \bibinfo{title}{Composite topological structure of domain walls in
		synthetic antiferromagnets}.
	\newblock \emph{\bibinfo{journal}{Sci. Rep.}} \textbf{\bibinfo{volume}{8}},
	\bibinfo{pages}{15794} (\bibinfo{year}{2018}).
	
	\bibitem{amin2023antiferromagnetic}
	\bibinfo{author}{Amin, O.} \emph{et~al.}
	\newblock \bibinfo{title}{Antiferromagnetic half-skyrmions electrically
		generated and controlled at room temperature}.
	\newblock \emph{\bibinfo{journal}{Nat. Nanotechnol.}} \bibinfo{pages}{1--5}
	(\bibinfo{year}{2023}).
	
	\bibitem{chmiel2018observation}
	\bibinfo{author}{Chmiel, F.~P.} \emph{et~al.}
	\newblock \bibinfo{title}{Observation of magnetic vortex pairs at room
		temperature in a planar $\alpha$-${F}e_{2}{O}_{3}/{C}o$ heterostructure}.
	\newblock \emph{\bibinfo{journal}{Nat. Mater.}} \textbf{\bibinfo{volume}{17}},
	\bibinfo{pages}{581--585} (\bibinfo{year}{2018}).
	
	\bibitem{jani2021antiferromagnetic}
	\bibinfo{author}{Jani, H.} \emph{et~al.}
	\newblock \bibinfo{title}{Antiferromagnetic half-skyrmions and bimerons at room
		temperature}.
	\newblock \emph{\bibinfo{journal}{Nature}} \textbf{\bibinfo{volume}{590}},
	\bibinfo{pages}{74} (\bibinfo{year}{2021}).
	
	\bibitem{radaelli2020micromagnetic}
	\bibinfo{author}{Radaelli, P.}, \bibinfo{author}{Radaelli, J.},
	\bibinfo{author}{Waterfield-Price, N.} \& \bibinfo{author}{Johnson, R.}
	\newblock \bibinfo{title}{Micromagnetic modeling and imaging of vortex| meron
		structures in an oxide| metal heterostructure}.
	\newblock \emph{\bibinfo{journal}{Phys. Rev. B}}
	\textbf{\bibinfo{volume}{101}}, \bibinfo{pages}{144420}
	(\bibinfo{year}{2020}).
	
	\bibitem{romming2013writing}
	\bibinfo{author}{Romming, N.} \emph{et~al.}
	\newblock \bibinfo{title}{Writing and deleting single magnetic skyrmions}.
	\newblock \emph{\bibinfo{journal}{Science}} \textbf{\bibinfo{volume}{341}},
	\bibinfo{pages}{636} (\bibinfo{year}{2013}).
	
	\bibitem{dupe2014tailoring}
	\bibinfo{author}{Dup{\'e}, B.}, \bibinfo{author}{Hoffmann, M.},
	\bibinfo{author}{Paillard, C.} \& \bibinfo{author}{Heinze, S.}
	\newblock \bibinfo{title}{Tailoring magnetic skyrmions in ultra-thin transition
		metal films}.
	\newblock \emph{\bibinfo{journal}{Nat. Commun.}} \textbf{\bibinfo{volume}{5}},
	\bibinfo{pages}{4030} (\bibinfo{year}{2014}).
	
	\bibitem{Simon2014}
	\bibinfo{author}{Simon, E.}, \bibinfo{author}{Palot{\'a}s, K.},
	\bibinfo{author}{R{\'o}zsa, L.}, \bibinfo{author}{Udvardi, L.} \&
	\bibinfo{author}{Szunyogh, L.}
	\newblock \bibinfo{title}{{ Formation of magnetic skyrmions with tunable
			properties in PdFe bilayer deposited on Ir (111)}}.
	\newblock \emph{\bibinfo{journal}{Phys. Rev. B}} \textbf{\bibinfo{volume}{90}},
	\bibinfo{pages}{094410} (\bibinfo{year}{2014}).
	
	\bibitem{crum2015perpendicular}
	\bibinfo{author}{Crum, D.~M.} \emph{et~al.}
	\newblock \bibinfo{title}{Perpendicular reading of single confined magnetic
		skyrmions}.
	\newblock \emph{\bibinfo{journal}{Nat. Commun.}} \textbf{\bibinfo{volume}{6}},
	\bibinfo{pages}{8541} (\bibinfo{year}{2015}).
	
	\bibitem{romming2015field}
	\bibinfo{author}{Romming, N.}, \bibinfo{author}{Kubetzka, A.},
	\bibinfo{author}{Hanneken, C.}, \bibinfo{author}{von Bergmann, K.} \&
	\bibinfo{author}{Wiesendanger, R.}
	\newblock \bibinfo{title}{Field-dependent size and shape of single magnetic
		skyrmions}.
	\newblock \emph{\bibinfo{journal}{Phys. Rev. Lett.}}
	\textbf{\bibinfo{volume}{114}}, \bibinfo{pages}{177203}
	(\bibinfo{year}{2015}).
	
	\bibitem{dos2016chirality}
	\bibinfo{author}{dos Santos~Dias, M.}, \bibinfo{author}{Bouaziz, J.},
	\bibinfo{author}{Bouhassoune, M.}, \bibinfo{author}{Bl{\"u}gel, S.} \&
	\bibinfo{author}{Lounis, S.}
	\newblock \bibinfo{title}{Chirality-driven orbital magnetic moments as a new
		probe for topological magnetic structures}.
	\newblock \emph{\bibinfo{journal}{Nat. Commun.}} \textbf{\bibinfo{volume}{7}},
	\bibinfo{pages}{13613} (\bibinfo{year}{2016}).
	
	\bibitem{fernandes2018universality}
	\bibinfo{author}{Fernandes, I.~L.}, \bibinfo{author}{Bouaziz, J.},
	\bibinfo{author}{Bl{\"u}gel, S.} \& \bibinfo{author}{Lounis, S.}
	\newblock \bibinfo{title}{Universality of defect-skyrmion interaction
		profiles}.
	\newblock \emph{\bibinfo{journal}{Nat. Commun.}} \textbf{\bibinfo{volume}{9}},
	\bibinfo{pages}{4395} (\bibinfo{year}{2018}).
	
	\bibitem{fernandes2020defect}
	\bibinfo{author}{Fernandes, I.~L.}, \bibinfo{author}{Bouhassoune, M.} \&
	\bibinfo{author}{Lounis, S.}
	\newblock \bibinfo{title}{Defect-implantation for the all-electrical detection
		of non-collinear spin-textures}.
	\newblock \emph{\bibinfo{journal}{Nat. Commun.}} \textbf{\bibinfo{volume}{11}},
	\bibinfo{pages}{1602} (\bibinfo{year}{2020}).
	
	\bibitem{Arjana2020}
	\bibinfo{author}{Arjana, I.~G.}, \bibinfo{author}{Lima~Fernandes, I.},
	\bibinfo{author}{Chico, J.} \& \bibinfo{author}{Lounis, S.}
	\newblock \bibinfo{title}{Sub-nanoscale atom-by-atom crafting of
		skyrmion-defect interaction profiles}.
	\newblock \emph{\bibinfo{journal}{Sci. Rep.}} \textbf{\bibinfo{volume}{10}},
	\bibinfo{pages}{14655} (\bibinfo{year}{2020}).
	
	\bibitem{bouhassoune2021friedel}
	\bibinfo{author}{Bouhassoune, M.} \& \bibinfo{author}{Lounis, S.}
	\newblock \bibinfo{title}{Friedel oscillations induced by magnetic skyrmions:
		From scattering properties to all-electrical detection}.
	\newblock \emph{\bibinfo{journal}{Nanomaterials}}
	\textbf{\bibinfo{volume}{11}}, \bibinfo{pages}{194} (\bibinfo{year}{2021}).
	
	\bibitem{lima2022spin}
	\bibinfo{author}{Lima~Fernandes, I.}, \bibinfo{author}{Bl{\"u}gel, S.} \&
	\bibinfo{author}{Lounis, S.}
	\newblock \bibinfo{title}{Spin-orbit enabled all-electrical readout of chiral
		spin-textures}.
	\newblock \emph{\bibinfo{journal}{Nat. Commun.}} \textbf{\bibinfo{volume}{13}},
	\bibinfo{pages}{1576} (\bibinfo{year}{2022}).
	
	\bibitem{muller2019spirit}
	\bibinfo{author}{M{\"u}ller, G.~P.} \emph{et~al.}
	\newblock \bibinfo{title}{Spirit: Multifunctional framework for atomistic spin
		simulations}.
	\newblock \emph{\bibinfo{journal}{Phys. Rev. B}} \textbf{\bibinfo{volume}{99}},
	\bibinfo{pages}{224414} (\bibinfo{year}{2019}).
	
	\bibitem{shinjo2000magnetic}
	\bibinfo{author}{Shinjo, T.}, \bibinfo{author}{Okuno, T.},
	\bibinfo{author}{Hassdorf, R.}, \bibinfo{author}{Shigeto, K.} \&
	\bibinfo{author}{Ono, T.}
	\newblock \bibinfo{title}{Magnetic vortex core observation in circular dots of
		permalloy}.
	\newblock \emph{\bibinfo{journal}{science}} \textbf{\bibinfo{volume}{289}},
	\bibinfo{pages}{930--932} (\bibinfo{year}{2000}).
	
	\bibitem{giannozzi2009quantum}
	\bibinfo{author}{Giannozzi, P.} \emph{et~al.}
	\newblock \bibinfo{title}{Quantum espresso: a modular and open-source software
		project for quantum simulations of materials}.
	\newblock \emph{\bibinfo{journal}{J. Phys.: Condens. Matter}}
	\textbf{\bibinfo{volume}{21}}, \bibinfo{pages}{395502}
	(\bibinfo{year}{2009}).
	
	\bibitem{dal2014pseudopotentials}
	\bibinfo{author}{Dal~Corso, A.}
	\newblock \bibinfo{title}{Pseudopotentials periodic table: From ${H}$ to
		${Pu}$}.
	\newblock \emph{\bibinfo{journal}{Comput. Mater. Sci.}}
	\textbf{\bibinfo{volume}{95}}, \bibinfo{pages}{337--350}
	(\bibinfo{year}{2014}).
	
	\bibitem{Papanikolaou2002}
	\bibinfo{author}{Papanikolaou, N.}, \bibinfo{author}{Zeller, R.} \&
	\bibinfo{author}{Dederichs, P.~H.}
	\newblock \bibinfo{title}{{Conceptual improvements of the KKR method}}.
	\newblock \emph{\bibinfo{journal}{J. Phys.: Condens. Matter}}
	\textbf{\bibinfo{volume}{14}}, \bibinfo{pages}{2799} (\bibinfo{year}{2002}).
	
	\bibitem{Bauer2014}
	\bibinfo{author}{Bauer, D. S.~G.}
	\newblock \emph{\bibinfo{title}{Development of a relativistic full-potential
			first-principles multiple scattering Green function method applied to complex
			magnetic textures of nano structures at surfaces}}
	(\bibinfo{publisher}{Forschungszentrum J{\"u}lich J{\"u}lich},
	\bibinfo{year}{2014}).
	
	\bibitem{Liechtenstein1987}
	\bibinfo{author}{Liechtenstein, A.}, \bibinfo{author}{Katsnelson, M.},
	\bibinfo{author}{Antropov, V.} \& \bibinfo{author}{Gubanov, V.}
	\newblock \bibinfo{title}{Local spin density functional approach to the theory
		of exchange interactions in ferromagnetic metals and alloys}.
	\newblock \emph{\bibinfo{journal}{J. Magn. Magn. Mater.}}
	\textbf{\bibinfo{volume}{67}}, \bibinfo{pages}{65 -- 74}
	(\bibinfo{year}{1987}).
	
	\bibitem{Ebert2009}
	\bibinfo{author}{Ebert, H.} \& \bibinfo{author}{Mankovsky, S.}
	\newblock \bibinfo{title}{Anisotropic exchange coupling in diluted magnetic
		semiconductors: \textit{Ab initio} spin-density functional theory}.
	\newblock \emph{\bibinfo{journal}{Phys. Rev. B}} \textbf{\bibinfo{volume}{79}},
	\bibinfo{pages}{045209} (\bibinfo{year}{2009}).
	
\end{thebibliography}
\end{document}